\pdfoutput=1
\documentclass[11pt]{article}

\usepackage[preprint]{acl}

\usepackage{times}
\usepackage{latexsym}

\usepackage[T1]{fontenc}
\usepackage[utf8]{inputenc}

\usepackage{microtype}

\usepackage{inconsolata}

\usepackage{graphicx}
\usepackage{hyperref}       % hyperlinks
\usepackage{url}            % simple URL typesetting
\usepackage{booktabs}       % professional-quality tables
\usepackage{amsfonts}       % blackboard math symbols
\usepackage{nicefrac}       % compact symbols for 1/2, etc.
\usepackage{microtype}      % microtypography
\usepackage{xcolor}         % colors

\usepackage{amsmath,amssymb,amsfonts}
\usepackage{algorithm,algorithmic}
\usepackage{tikz}
\usepackage{graphicx}
\usepackage{textcomp}
\usepackage{wrapfig}
\usepackage{subfigure}
\usepackage{times}
\usepackage{epsfig}
\usepackage{todonotes}
\usepackage{comment}
\usepackage{multicol,multirow}
\usepackage{adjustbox}
\usepackage{tabularx}
\usepackage{booktabs}
\usepackage{lipsum}
\usepackage{url}
\definecolor{warningcolor}{RGB}{255, 0, 0}

\title{DePrompt: Desensitization and Evaluation Personal Identifiable Information in Large Language Model Prompts}

\author{Xiongtao Sun\textsuperscript{1}, Gan Liu\textsuperscript{2}, 
Zhipeng He\textsuperscript{1}, 
Hui Li\textsuperscript{1}, Xiaoguang Li\textsuperscript{1}, \\
\textsuperscript{1}Xidian University,~\textsuperscript{2} Hainan Universit\\
\texttt{xtsun@stu.xidian.edu.cn}\\
}

\begin{document}
\maketitle
\begin{abstract}
Prompt serves as a crucial link in interacting with large language models, widely impacting the accuracy and interpretability of model outputs. However, acquiring accurate and high-quality responses necessitates precise prompts, which inevitably pose significant risks of personal identifiable information (PII) leakage. Therefore, this paper proposes DePrompt, a desensitization protection and effectiveness evaluation framework for prompt, enabling users to safely and transparently utilize large language models. Specifically, by leveraging large model fine-tuning techniques as the underlying privacy protection method, we integrate contextual attributes to define privacy types, achieving high-precision PII entity identification. Additionally, through the analysis of key features in prompt desensitization scenarios, we devise adversarial generative desensitization methods that retain important semantic content while disrupting the link between identifiers and privacy attributes. Furthermore, we present utility evaluation metrics for prompt to better gauge and balance privacy and usability. Our framework is adaptable to prompts and can be extended to text usability-dependent scenarios. Through comparison with benchmarks and other model methods, experimental evaluations demonstrate that our desensitized prompt exhibit superior privacy protection utility and model inference results.
\end{abstract}

% !TEX root = main.tex

\vspace*{-1em}

\section{Introduction}\label{intro}

\begin{figure}[t]
\centering
\includegraphics[width = 7.5cm]{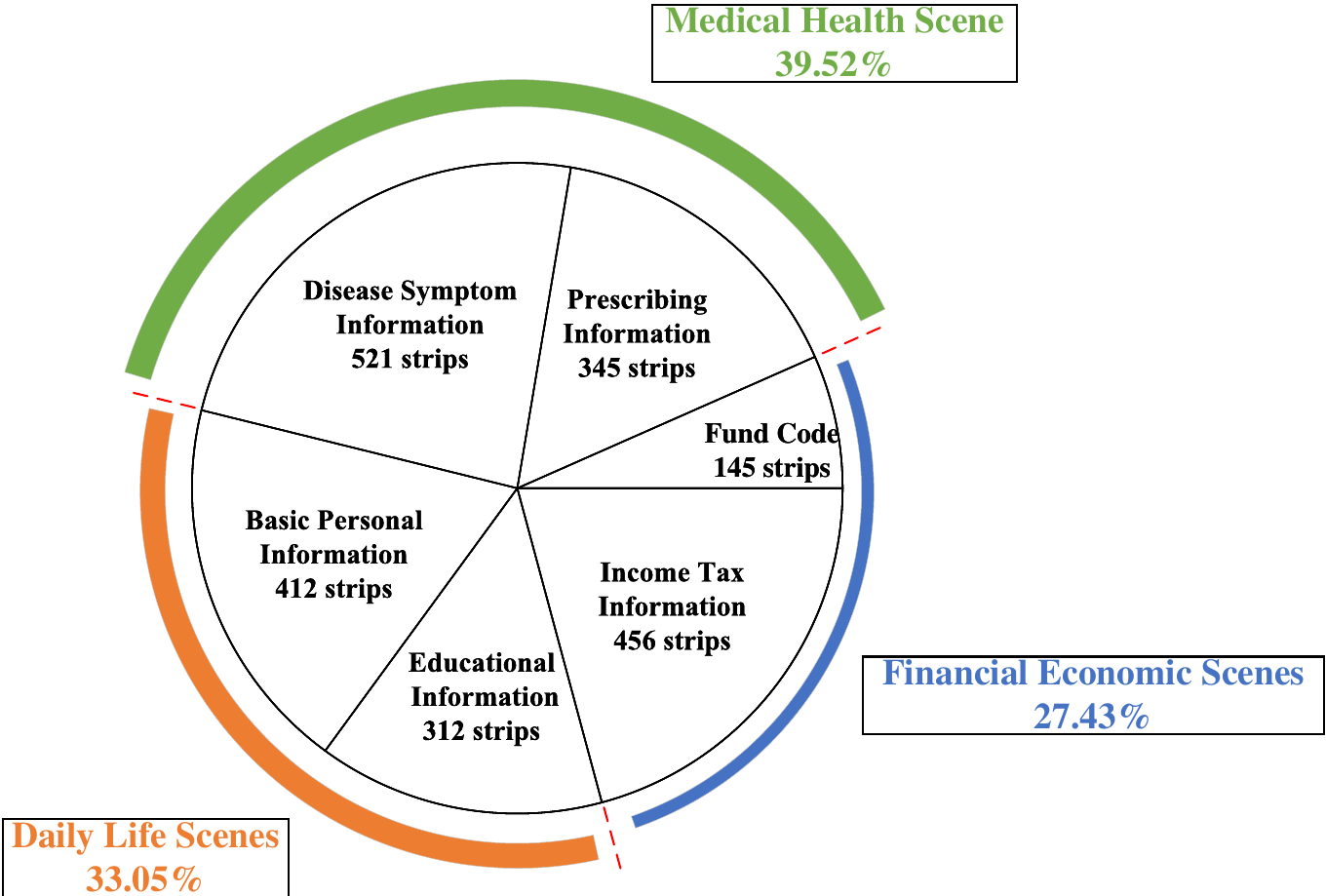}
\caption{Public prompt dataset PII leakage situation.}
\label{fig:intro}
\end{figure}

The emergence of Large Language Models (LLMs) such as GPT-4~\cite{gpt3} has profoundly transformed the way we interact with technology. This transformation spans from intelligent assistants to customized content generation, showcasing their significant impact on our everyday lives and the field of natural language processing (NLP)~\cite{nlp}. Specifically, the role of prompts in fine-tuning~\cite{lester2021power} and applying LLMs is crucial. Prompts provide context and task guidance to the models, directing them to learn specific tasks and contexts, thereby influencing the way the models learn from data and generate results~\cite{prompt1}. To enhance the quality of interaction with LLMs, whether in prompt tuning during the model training phase or prompt-based generation during the model inference phase, it is necessary to upload a large number of prompt samples for exploration and evaluation.

However, prompts, as the link between users and LLMs, also bring about security and privacy risks. In security-sensitive scenarios, prompt-related security vulnerabilities mainly revolve around robustness~\cite{jailbreaking1,injection1,injection2} and privacy. While robustness can pose privacy threats~\cite{jailbreaking2}, this study focuses directly on researching privacy issues. Prompts may involve users’ Personally Identifiable Information (PII) or sensitive information of organizations, directly leading to privacy breaches~\cite{promptprivacy1}. Furthermore, LLMs have been shown to have the capability to memorize data through prompts, thereby resulting in indirect privacy leaks~\cite{promptprivacy2}. Privacy breaches can have serious consequences, especially in the case of PII leaks, which could lead to financial losses, susceptibility to social engineering attacks, and even identity theft. As illustrated in Figure~\ref{fig:intro}, we extracted 1000 prompts from publicly available prompt datasets, from which a significant number of different types of PII entities were identified. Therefore, there is an urgent need to develop privacy-preserving prompt schemes for LLMs.

\textbf{Our Distinction from Previous Research.} To tackle the aforementioned challenges, prior work has proposed various privacy-preserving prompt schemes based on Privacy-Preserving Data Publishing (PPDP) and Natural Language Processing (NLP) technologies. Scheme~\cite{hong2024dpopt} was the first to combine differential privacy with prompt fine-tuning. It introduced transferable discrete prompts through an exponential differential privacy mechanism, thereby achieving the generation of privacy-preserving prompts. Scheme~\cite{yu2024privacy} focused on the annotation and alignment stages of prompt, utilizing a private fine-tuning generator to produce synthetic instructions, ensuring differential privacy and thereby achieving privacy protection for prompts. Differential privacy methods focus on protecting the statistical characteristics of an entire prompt dataset in a fine-tuning scenario, rather than during the inference phase of prompt applications. Consequently, they are not suitable for protecting PII in individual prompt instances during dynamic application, thus unable to address direct privacy leaks within prompts. Additionally, the introduced noise inevitably leads to grammatical errors and semantic biases, thereby reducing the usability of prompts.

The approaches~\cite{DBLP:conf/iccsa/CampanileBMMRV50, DBLP:journals/jbi/MeystreFFSSS14, liu2023gptdeid} utilize common NLP models such as LSTM, BiLSTM-CRF, and GPT-4 for text privacy recognition. With advancements in NLP, these methods have achieved outstanding accuracy in entity recognition. However, post entity recognition, they often resort to traditional desensitization methods such as deletion and masking, leading to unacceptable losses in prompt usability. Moreover, dynamic changes in privacy definitions necessitate re-annotating data and retraining models. This introduces ongoing costs on one hand and inadvertently increases the risk of privacy breaches on the other.

\textbf{Our Approach.} Therefore, to address the protection of PII during the inference stage of LLM applications, we propose an efficient prompt privacy desensitization and utility assessment framework, \textbf{DePrompt}. Specifically, we introduce key elements for prompt privacy protection, leading to the design of a dynamic adversarial generative desensitization method, achieving a better balance between privacy and usability. Finally, we present metrics for evaluating prompt privacy and usability, enhancing the interpretability of the framework.

\textbf{Contributions.} Our contributions can be summarized as follows:

\begin{enumerate}
\item \textbf{DePrompt achieved high-accuracy recognition of PII entities.} In our framework, we analyzed common scenarios for prompts and defined dynamic privacy concept templates. By combining LLMs fine-tuning techniques, we achieved efficient PII recognition, with an average accuracy of 95.95\%, providing an effective foundation for subsequent desensitization.

\item \textbf{Our framework proposed an efficient adversarial generative desensitization method.} We outlined key attributes essential for protecting PII in prompts. DePrompt, recognizing the differential impact of various entities on both semantics and privacy, harnesses the robust text comprehension and generation capabilities of LLMs to develop an adversarial generative desensitizationmethod. This approach enables a more effective balance between privacy and utility.

\item \textbf{Our framework introduced a comprehensive and rational utility metric for prompt desensitization.} Considering both privacy and utility, we comprehensively measured the desensitization effectiveness of prompts by employing various levels of privacy attacks as well as assessing semantic loss, differences in model responses, and readability. Furthermore, through comparisons with traditional anonymization  methods, we demonstrated the high practicality and balance of our framework.
\end{enumerate}
\section{Related works}
\label{realted}
In this section, we introduce some existing methods for prompt privacy anonymization, including PPDP and NLP, as well as some standard metrics for evaluating the protected text.

\subsection{Privacy-preserving data publishing}
PPDP is a privacy-enhancing technology designed to facilitate the public usage of data while preventing the disclosure of sensitive information ~\cite{gehrke2006models}, such as K-anonymity ~\cite{k-anonymity} and differential privacy \cite{dwork2006differential}. 
Hong \textit{et al.} ~\cite{hong2024dpopt} proposed DP-OPT, the first private prompt generation mechanism. By transferring prompts through the Deep Language Network (DLN) and combining exponential differential privacy, they achieved competitive performance in generating private prompts.
Similarly, Yu \textit{et al.} ~\cite{yu2024privacy} combined differential privacy with prompt tuning, with the additional focus on the annotation phase of prompt tuning. Therefore, they trained a private fine-tuning generator for prompt generation. 

The above method only applies to structured datasets and cannot be tailored to individual prompt application scenarios. However, for prompt, semantics are more important than statistical distributions. Therefore, methods such as deletion, substitution, and generalization are still commonly used to protect personally identifiable information (PII) in textual content ~\cite{mamede2016automated}. However, these methods often result in unacceptable utility loss of the data. Therefore, it is necessary to design a PPDP method suitable for individual prompts while maintaining a certain level of usability.

\subsection{Natural language processing}
NLP methods have not yet been applied to prompt privacy protection, primarily focusing on tasks such as detecting identifiers and privacy attributes.
Liu \textit{et al.} ~\cite{DBLP:conf/aims2/LiuLLDTX19} designed a novel capsule-LSTM network to better capture entity-related information conveyed in clinical text, combining the strong expressive power of capsule networks with the sequential modeling capability of LSTM networks. 
Campanile \textit{et al.} ~\cite{DBLP:conf/iccsa/CampanileBMMRV50} proposed an automatic clinical information de-identification method. This method first conducts advanced assessment of the amount and format of clinical record information, and subsequently performs detailed research on the overlap between selected clinical information types and protected health information (PHI). 
Meystre \textit{et al.} ~\cite{DBLP:journals/jbi/MeystreFFSSS14} proposed an automated solution using the BiLSTM-CRF model to remove private information from clinical records, effectively identifying protected health information from unstructured text. 
Liu \textit{et al.} ~\cite{liu2023gptdeid} proposed the utilization of GPT-4 for medical text data processing and de-identification, though this implies the uploading of private data to a third party.

After NLP recognition, entities are often assumed to be deleted or replaced, resulting in unnecessary loss of data utility. The definition of privacy is limited to predefined entity categories. When there is a need to dynamically change the privacy definition based on the context, data annotation and model training need to be re-performed, and privacy leakage risks still exist. Therefore, there is a need to dynamically define the concept of privacy, and to use PPDP methods appropriately after identifying privacy entities.

\subsection{Evaluation metrics}
Ni \textit{et al.} ~\cite{ni2022data} proposed a data anonymization evaluation framework for datasets in Internet of Things (IoT) scenarios. The framework assesses commonly used data anonymization algorithms based on privacy protection level, data utility, and performance. 
Pil{\'a}n \textit{et al.} ~\cite{pilan2022textbenchmark} proposed a novel benchmark and relevant evaluation metrics to assess the performance of text anonymization methods in the field of NLP. Additionally, they provided a new open-source annotated corpus. 
Benet \textit{et al.} ~\cite{manzanares2022automatic} proposed a re-identification attack on anonymous text, using state-of-the-art deep learning language models combined with background knowledge available to potential attackers, to provide an authentic assessment of risk disclosure.

In privacy measurement, existing evaluation metrics, such as single IR-based metrics, do not consider the relationship between privacy attributes and identifiers and the varying impact of different entities on the text. In utility measurement, most metrics focus on the overall dataset and cannot be tailored to individual data. Therefore, the existing evaluation metrics system is not applicable to prompt.

\section{System and threat models}
\label{system}
In this section, we first define the system and threat models of our framework.
\subsection{System model}
\begin{figure*}
\centering
\includegraphics[width = 1.8\columnwidth]{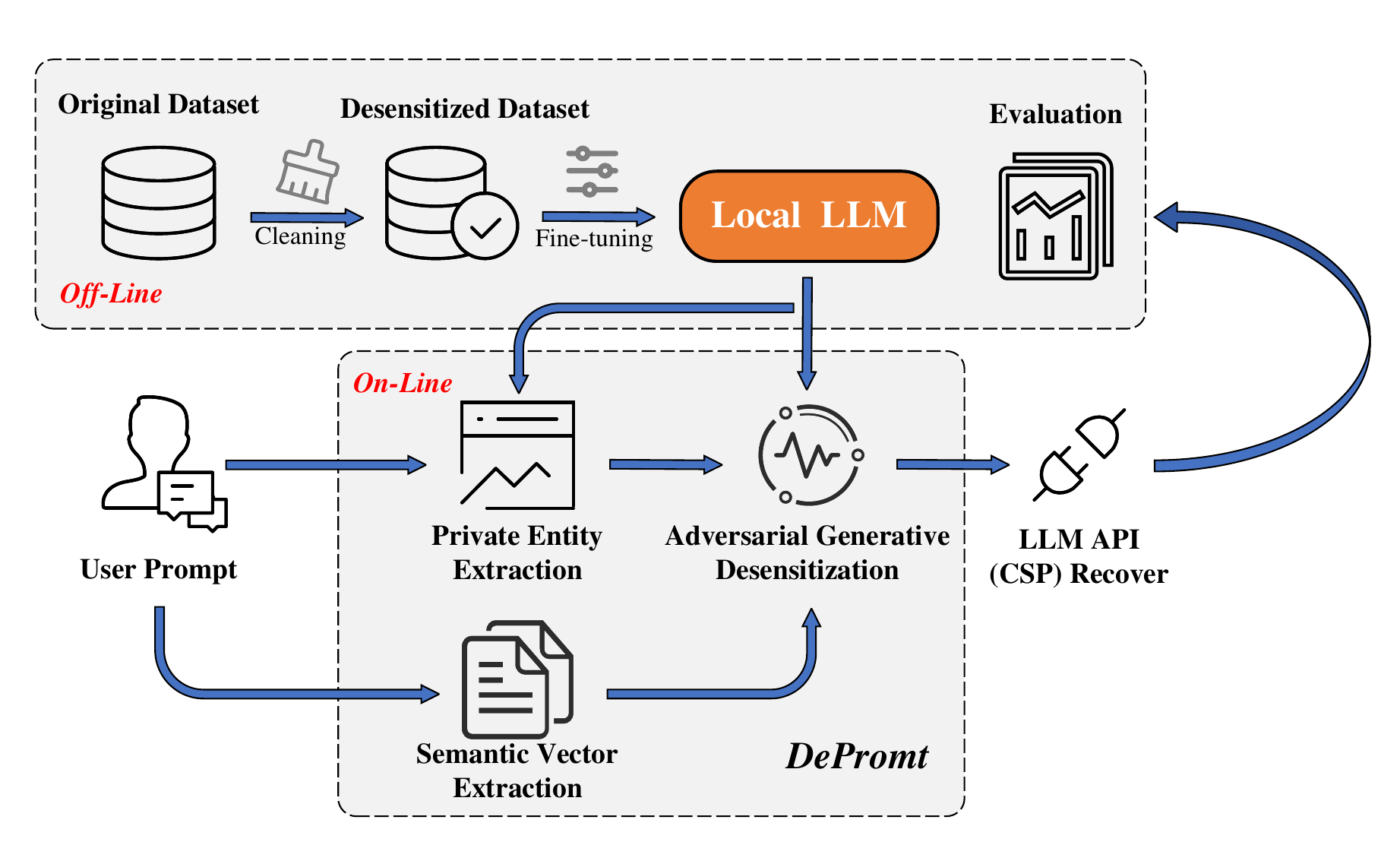}
\caption{The system model of our framework.}
\label{fig:system}
\end{figure*}

Our frameworl concentrates on the protection of PII in prompt, which consists of three entities, including a LLM user $U$ , local protection mechanism for prompt \textit{DePrompt}, and a LLM cloud service provider \textit{CSP}, as shown in \figurename~\ref{fig:system}. The commonly used symbols and corresponding definitions are listed in \tablename~\ref{TabSymbols}.

\textbf{LLM User}.The LLM user provides prompts that can clearly and accurately express their intentions, and uploads them to the \textit{CSP} for LLM, in order to obtain the generated inference results of the LLM.

\textbf{DePrompt}.The main purpose of \textit{DePrompt} is to perform PII identification and anonymization on user-submitted prompts. This process involves semantic vector extraction, private entity extraction, adversarial generative desensitization, and measurement of privacy and usability.

\textbf{CSP}. The LLM cloud service provider possess abundant computing and communication resources to offer interfaces for large language models. They are responsible for receiving user prompts and delivering the inference results of the large language model.

\begin{table*}
\renewcommand\arraystretch{1.3}
\center
\small
\caption{{Symbols of our scheme}}
\label{TabSymbols}
\setlength\tabcolsep{1.3pt}
\renewcommand\arraystretch{1.3}
\begin{tabular}{ll}
 \toprule
 \textbf{Symbols}  & \textbf{Definition} \\ \hline
 $p$ & U’s original prompt. \\
 $p_n$ & U's prompt dataset. \\
 $p_{de}$ & Depromt desensitized prompt. \\
 $S$ & Semantic entity vector. \\
 $s_t$ & The $t$-th entity in the semantic vector. \\
 $K$ & The length of the semantic entity vector. \\ 
 $\mathcal{L}$ & Privacy entity vector. \\
 $l_t$ & The $t$-th entity in the privacy vector. \\
 $d_t$ & The $t$-th direct identifier. \\
 $q_t$ & The $t$-th quasi identifier. \\
 $a_t$ & The $t$-th confidential attribute. \\
 $M$ & The length of the privacy entity vector.  \\
 $N$ & The length of the adversarial generation prompt vector.  \\
 % {$c_t^n$} & The comparison value of ${S}_t^n$. \\
 % $a_t^n, b_t^n$ & Random numbers for perturbing $c_t^n$. \\
 % $V^n$ & A vector denoting the index of the queried leaf. \\
 % $\widetilde{c}^n_t, \widetilde{V}^n$ & Perturbed $c_t^n$ and $V^n$. \\
 % {$\widetilde{L}^n_Q$} & The queried leaf on perturbed $n$-th tree. \\
 $R_p$ & The prompt inference result. \\
 $\mathcal{L} \leftarrow \mathbb{R}(p_{de})$ & Extract the set $\mathcal{L}$ of all PII sequences in $p_{de}$.\\
 $d' \leftarrow \mathbb{I}(p_{de},a,D)$ & Infer $p_{de}$ which identifier $d'$  belongs to from $D$ with PII $a$. \\
 \bottomrule
\end{tabular}
\end{table*}

\subsection{Threat model}
\begin{figure}
\centering
\includegraphics[width = 0.9\columnwidth]{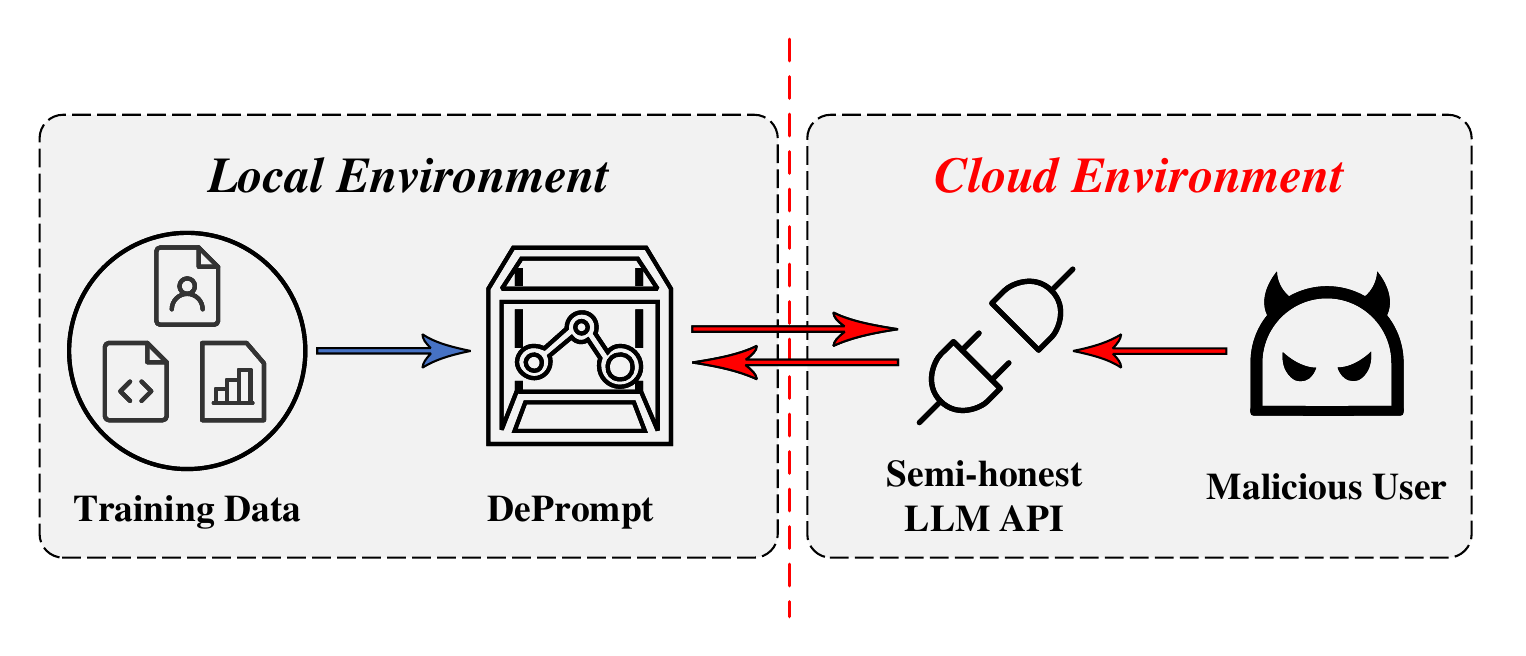}
\caption{Prompt workflows and sources of threats.}
\label{fig:threat}
\end{figure}

As shown in the \figurename~\ref{fig:threat}, in our threat model, \textit{CSP} is considered honest-but-curious \cite{goldreich2019play} and that the communication channel between local and cloud environments is susceptible to monitoring. In other words, Adversaries have access to the prompts submitted by $U$ to the LLM \textit{CSP}, as well as the LLM frameworks and interfaces used. However, they do not maliciously modify the prompts or the feedback results from the models. For a given model, we assumes that adversaries utilize white-box background knowledge for extracting and analyzing personal sensitive information, which is then applied to downstream tasks such as user profiling, thereby infringing upon individual privacy.

\section{Preliminaries}
\label{prelimi}
This section begins with an introduction to fundamental concepts of data anonymization. Building on these concepts, we identify three critical attributes in Prompt-based data anonymization: semantic relevance, subject linkage, and entropy uncertainty, each essential for subsequent analysis. Additionally, we demonstrate the importance of LLM fine-tuning techniques as a significant tool for implementing attribute-based thinking within our framework.

\subsection{Anonymization Key Terms}
The current legal landscape, including legislations like the European General Data Protection Regulation (GDPR), the California Consumer Privacy Act (CCPA) in the United States, and China's Personal Information Protection Law (PIPL), typically mandates organizations to obtain explicit consent when collecting and processing personal data. Moreover, these regulations require appropriate measures such as encryption or anonymization of stored personal data to mitigate risks of data breaches and misuse. This underscores the significance of anonymization efforts, particularly in ensuring the security of personal data and complying with privacy laws. The following are some key terms pertaining to anonymization\cite{2021sokanonymisation}:

\begin{itemize}
    \item \textbf{Direct Identifier} refers to information that contains unique values capable of uniquely identifying an individual's identity. Examples of such information include ID card numbers, passport numbers, driver's license numbers, and more.
    
    \item \textbf{Quasi Identifier} refers to indirectly identifiable information, while not capable of uniquely identifying an individual, can assist in confirming the identity of a specific person when combined with other information. There is no fixed type or scope for this information. Examples include gender, nationality, occupation, and more.
    
    \item \textbf{Confidential Attribute} refers to highly sensitive and confidential information, the disclosure of which could have a significant impact on individuals. Examples include personal health conditions, religious beliefs, political views, and more.

    \item \textbf{Personally Identifiable Information} refers to all information that can be used to identify, contact, or locate an individual. In this paper, we consider it as the union of Direct Identifiers, Quasi Identifiers, and Confidential Attributes.
        
\end{itemize}

\subsection{Prompt-based anonymization attributes}
\label{attributes}
Based on the requirement of anonymity and the characteristics of large language models, we propose three important features for PII in prompt: \textbf{linkability}\cite{kim2023propile}, \textbf{semanticity}, and \textbf{uncertainty}. These features are incorporated into the DePrompt anonymization method, which is integrated throughout the entire anonymization framework.

\subsubsection{Linkability}
In using prompt, from the perspective of privacy leakage, the disclosure of privacy attributes alone may not necessarily pose significant risks. For example, when LLMs are employed in disease diagnosis, the symptom information contained in prompt cannot be directly linked to specific patients. Therefore, the crucial aspect of desensitization is to break the linkability between privacy attributes and identifiers. 

\textbf{Definition 1 Linkability in prompt.} Let $\mathcal{L}:=\left\{d_{1}, \ldots,q_{n-1},q_{n}, \ldots,a_{M}\right\}$ be M PII items relevant to a subject $U$. Each $d_{m}$ represents direct identifier, $q_{m}$ represents quasi identifier, and $a_{m}$ represents confidential attribute. Direct linkability refers to the scenario where $a_{m}$ can be associated with  specific individuals $U$ through specific $d_{m}$. On the other hand, Indirect linkability refers to the scenario where $a_{m}$ can be linked to specific individuals $U$ with a certain probability through specific $q_{m}$.

\subsubsection{Semanticity}
During the process of prompt desensitization, it is crucial to consider the correlation between different PII and semantics, as it determines the appropriate desensitization methods to be employed. In cases where there is little or weak correlation, methods such as deletion or generalization can be directly applied for data desensitization. However, if there is a strong correlation, desensitization methods with minimal semantic loss need to be utilized.

\textbf{Definition 1 Semanticity in prompt.} Let $S = \left\{s_1, s_2,\ldots, s_K\right\}$ be K semantic keywords and  $\mathcal{L}:=\left\{d_{1}, \ldots,q_{n-1},q_{n}, \ldots,a_{M}\right\}$ be M PII items from a prompt $p$. If a singular entity appears in both $S$ and $\mathcal{L}$, we refer to this entity as having privacy semantic relevance.

\subsubsection{Uncertainty}
The utilization of desensitization techniques such as generalization, K-anonymity, and differential privacy is fundamentally aimed at increasing the uncertainty of PII. This process can be understood as augmenting the information entropy of the data. By reducing data specificity, desensitization methods enhance the level of confusion in the data, thus fortifying the protection of individual identities and achieving the objective of privacy preservation.

\textbf{Definition 3 Uncertainty in prompt.} The $\mathcal{L}:=\left\{d_{1}, \ldots,q_{n-1},q_{n}, \ldots,a_{M}\right\}$ represents PII vectors extracted from the prompt $p$. Uncertainty refers to the process of obfuscating confidential attributes $a_{m}$, thereby preventing the precise extraction of private information from $p$.
\section{DePrompt}
\label{deprompt}
In this section, we first present our proposed framework DePrompt with an overview. Then the detailed processes and steps of DePrompt are introduced.
\subsection{Overview}

As mentioned above, prior to uploading to the \textit{CSP}, the original prompt requires anonymization to mitigate the risk of privacy breaches. Unlike traditional data anonymization scenarios, anonymizing prompts necessitates ensuring their usability and precludes the use of conventional anonymization methods applicable to datasets, such as differential privacy or k-anonymity. Therefore, we propose DePrompt, a prompt anonymization framework deployable locally. The aim is to achieve the desensitization protection of PII within prompts while ensuring a certain level of prompt usability. Within DePrompt, we consider semanticity, linkability, and uncertainty to achieve a better balance between privacy and usability. 

\begin{figure*}
\centering
\includegraphics[width = 1.8\columnwidth]{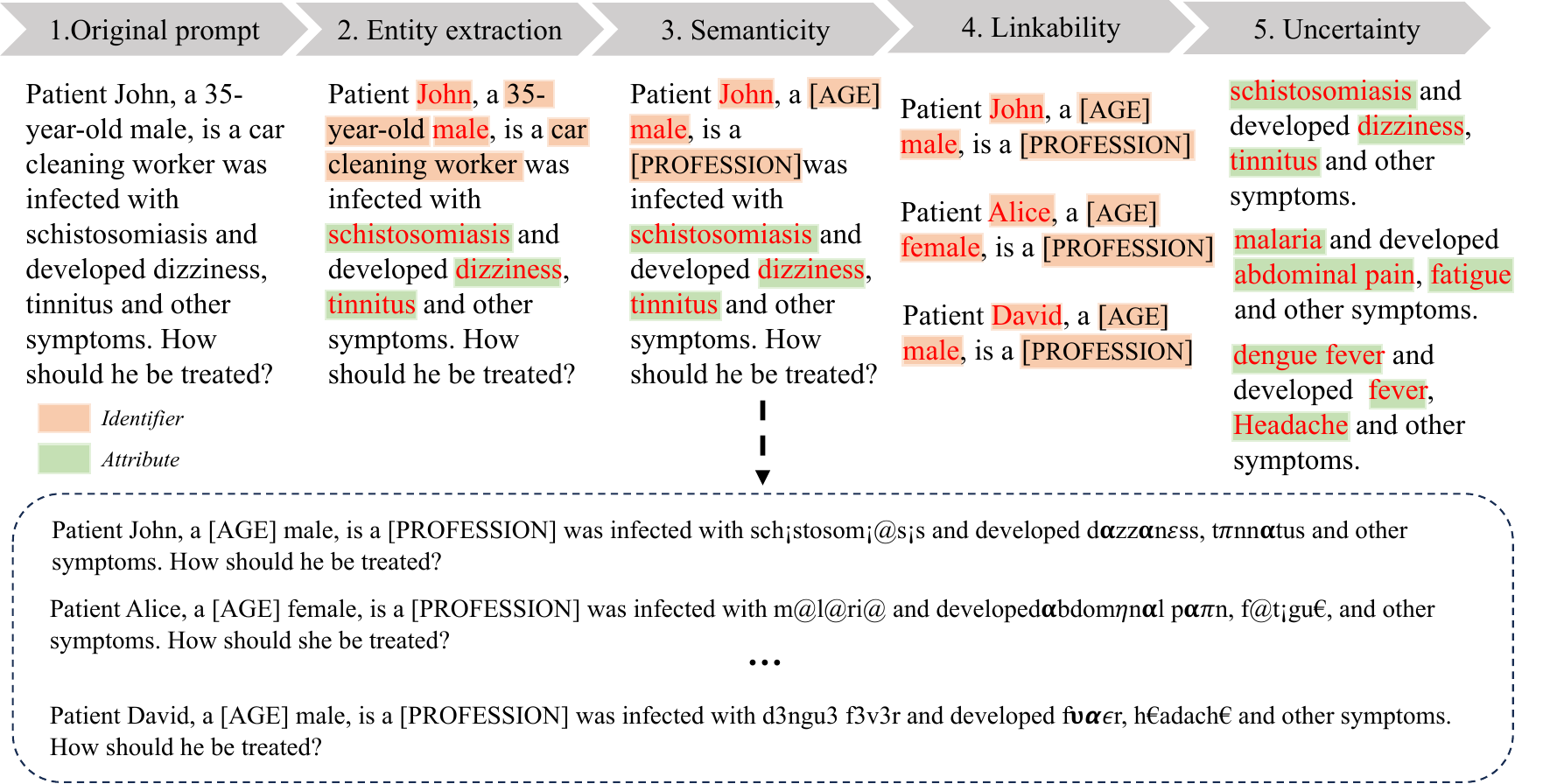}
\caption{An example prompt in the complete workflow of DePrompt}
\label{fig:process}
\end{figure*}

\subsection{Construction of DePrompt}
DePrompt primarily consists of pre-deployment fine-tuning of large language models and post-deployment semantic and privacy entity extraction, as well as adversarial generative desensitization. In the \figurename~\ref{fig:process}, we illustrate the complete process of a prompt in DePrompt.

\subsubsection{Large language model fine-tuning}
\label{fine-tuning}
Since DePrompt includes tasks such as scene recognition, privacy entity identification, and adversarial text generation, the model needs to possess the ability to comprehend complex scenarios and rich contextual information. It should also handle semantic associations and have the capacity to capture logical coherence between sentences, ensuring contextual consistency while generating adversarial text that is both perplexing and realistic. Therefore, we utilize LLM fine-tuning techniques to implement the model base for DePrompt, following the steps below:

\paragraph{Step 1. Data Collection.}

From several publicly available prompt datasets, we gathered prompts with distinct scene characteristics. These were categorized into medical, daily, and financial scenes, and manually annotated for privacy entities.

\paragraph{Step 2. Prompt Design.}

\begin{figure}
\centering
\includegraphics[width = \columnwidth]{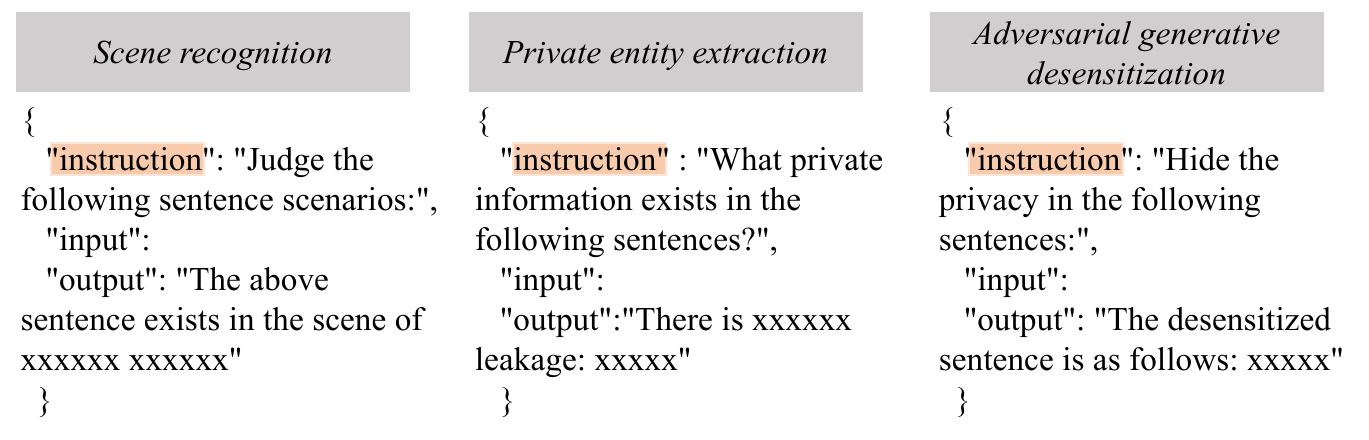}
\caption{Specific formatting of prompts in fine-tuning.}
\label{fig:prompts}
\end{figure}

In light of the downstream tasks involved in DePrompt, it is necessary to design different prompts for subsequent fine-tuning. We extract the inputs from the collected prompts in order to restructure them. The specific formatting is illustrated in the \figurename~\ref{fig:prompts}, with the top portion providing an example and the bottom portion outlining the precise format requirements.

\paragraph{Step 3. LORA Fine-tuning.}

In this step, firstly, download the weights of the large model. Then, convert the downloaded model into the HuggingFace format and expand the vocabulary. Finally, set the required number of layers for fine-tuning and utilize LORA for model training.

\subsubsection{Entity extraction}
\label{extraction}

\paragraph{Step 1. Semantic entity extraction.}

In this step, we utilize the TextRank algorithm \cite{DBLP:conf/emnlp/MihalceaT04} to construct a text as an undirected graph model. We employ graph ranking algorithms to calculate the importance of nodes, thereby extracting the top $K$ keywords. This process enables us to obtain semantic entity vectors $S = \left\{s_1, s_2,\ldots, s_K\right\}$, facilitating the analysis and comprehension of the semantic significance of the text.

\paragraph{Step 2. Privacy definition.}

In DePrompt, we consider scenarios where large models are commonly used and where privacy requirements are high, encompassing daily, medical, and financial contexts. We have defined the following types of privacy entities:

\textbf{PERSON} Personal statistical information, including name, age, gender, nationality, occupation, address, and employer.

\textbf{CODE} Identification codes, including ID card numbers, employee ID numbers, residency permit numbers, and passport numbers.

\textbf{CONTACT} Personal contact information, including personal phone numbers, email addresses, account details, and associated information.

\textbf{HEALTH} Health status data, including medical history, allergy records, symptoms, and health examination results.

\textbf{MEDICAL} Medical application data, including diagnostic outcomes, medication details, surgical records, and inpatient records.

\textbf{PAYMENT} Payment data, including transaction records, expenditure amounts, and insurance information.

\textbf{ASSET} Financial asset-related information, including personal asset data, bank branch numbers, service point identifiers, fund codes, and securities codes.

\paragraph{Step 3. Private entity extraction.}

In this phase, utilizing the locally fine-tuned large-scale model, we perform context recognition and privacy entity extraction based on the privacy definitions across various scenarios, resulting in privacy entity vectors $\mathcal{L}:=\left\{d_{1}, \ldots,q_{n-1},q_{n}, \ldots,a_{M}\right\}$.

\subsubsection{Adversarial generative desensitization}
As mentioned in Section~\ref{attributes}, the aspects of semantics, linkability, and uncertainty are crucial in anonymous process. Leveraging the powerful text generation capabilities of LLM, we introduce adversarial generative anonymization as outlined in Algorithm~\ref{generative}.

\begin{algorithm}[!ht]
\caption{Adversarial generative desensitization}
\label{generative}
\textbf{Input (U):} $p$ \\
\textbf{Input (Extract):} $S, \mathcal{L}$ \\
\textbf{Output:} $p_{de}$
\begin{algorithmic}[1]
% \LCOMMENT{Semanticity}

\FOR{$n = 1, \cdots, M$}
\IF{$l_n \notin S$}
\STATE{{$Scrub(p,l_n)$}}
\ENDIF
\ENDFOR

% \LCOMMENT{Linkability}
\FOR{$d,q \in \mathcal{L}, S$}
\STATE{{$d_{t'},q_{t'} \leftarrow Generative(d_t,q_t)$}}
\ENDFOR

% \LCOMMENT{Uncertainty}
\FOR{$n = 1, \cdots, N$}
\IF{$n \ne N$}
\FOR{$a \in \mathcal{L}, S$}
\STATE{{$a_{t'} \leftarrow Generative(a_t)$}}
\STATE{{$a_{t''} \leftarrow Adversial(a_{t'})$}}
\STATE{{$p_{n} \leftarrow Replace(p,d',q',a')$}}
\ENDFOR
\ENDIF
\FOR{$a \in \mathcal{L}, S$}
\STATE{{$a_{t''} \leftarrow Adversial(a_{t'})$}}
\STATE{{$p_{N} \leftarrow Replace(p,d',q',a')$}}
\ENDFOR
\ENDFOR
\STATE{{$p_{de} = Shuffle\left\{p_1, p_2,\ldots, p_N\right\}$}}
\RETURN{$p_{de}$}
\end{algorithmic}
\end{algorithm}

\paragraph{Semanticity(Line 1-5).} After receiving the prompt query $p$ from $U$, Locally extract $S$ and $\mathcal{L}$. We iterate through all the entities in $\mathcal{L}$, and for semantically unrelated entities, traditional $Scrub$ techniques such as deletion and masking can be employed to protect them.

\paragraph{Linkability(Line 6-8).} For semantically linked identifiers $d$ and $q$, we leverage the powerful generative capabilities of  fine-tuned LLM for $Generative$ replacement. Imitating the prompt pattern for generation replacement, for instance, using “Alice” instead of “Jack”. This generation disrupts the linkage between $U$ and their private attributes $a_m$.

\paragraph{Uncertainty(Line 9-23).} For semantically linked private attributes $a_m$, we initially employ a generative approach. Subsequently, we utilize $Adversarial$ techniques to perturb these attributes using various special characters, rendering traditional Named Entity Recognition (NER) techniques incapable of automatically extracting the private attributes, while still enabling LLMs to comprehend and infer. We replace $d$, $q$, and $a$ within the submitted prompt $p$ by user $U$, and iterate the aforementioned process $N-1$ times. we concatenate a prompt that has been subjected to adversarial perturbation, containing the original private attributes, resulting in an N-length prompt vector $p_{de}$, followed by a $Shuffle$ operation. Finally, submit multiple prompt to LLM CSP to confuse the malicious attacker and get a LLM inference response.

\section{Anonymization Evaluation}
\label{evalu}
In this section, We develope a set of metric standards for privacy and usability concerning prompt input and large model inference output.

\subsection{Privacy metircs}
In the privacy measurement task, the approach outlined in research \cite{promptprivacy2} can be leveraged by employing advanced attacks on the prompt. Through the implementation of privacy attacks, potential privacy risks can be exposed, and the effectiveness of the adopted privacy protection measures can be assessed. In this paper, our primary focus lies in the utilization of attacks for PII extraction and identifier linkage.

\begin{algorithm}[h]
\caption{PII Extraction}
\label{piiextract}
\textbf{Input:} $p^n,\mathbb{R}$ \\
\textbf{Output:} $\varepsilon_{e}$
\begin{algorithmic}[1]
\FOR{$n = 1, \cdots, T$}
\STATE{{$p \leftarrow RandomSelect(p^n)$}}
\STATE{{$E \leftarrow HumanExtract(p)$}}
\STATE{{$p_{de} \leftarrow Desensitization(p)$}}
\STATE{{$\mathcal{L} \leftarrow \mathbb{R}(p_{de})$}}
\STATE{{$\varepsilon \leftarrow \varepsilon+\mathbb{E}[\frac{\|E \cap \mathcal{L}\|}{\|E\|}]$}}
\ENDFOR
\RETURN{$\varepsilon/T$}\\
\end{algorithmic}
\end{algorithm}

\textbf{PII Extraction.} In PII extraction attack, the attacker's goal is to extract $U$’s PII  from the prompt as comprehensively as possible.

As illustrated in Algorithm \ref{piiextract}, given a user prompt dataset $p^n$ and an extraction algorithm $\mathbb{R}$, the process commences by $RandomSelect$ a prompt $p$ from the dataset. Subsequently, the private vector $E$ is $HumanExtract$ from this prompt. Then, the prompt undergoes $Desensitization$ to achieve a desensitized prompt $p_{de}$. Next, the extraction algorithm $\mathbb{R}$ is employed to obtain the private vector $\mathcal{L}$. Finally, the two private vectors are compared and their intersection is computed as $\varepsilon$. This entire process is iterated $T$ times, and the average value $\varepsilon_{e}$ is returned.

\textbf{Identifier Linkage.} In identifier linkage attack, the adversary goal is to associate PII with a specIFic $U$. It is assumed that private attributes are leaked and that the adversary possesses a list of direct identifiers.

\begin{algorithm}[h]
\caption{Identifier Linkage}
\label{idenlink}
\textbf{Input:} $p^n, \mathbb{I}$ \\
\textbf{Output:} $\varepsilon_{i}$
\begin{algorithmic}[1]
\STATE{{$D \leftarrow HumanExtract(p^n)$}}
\FOR{$n = 1, \cdots, T$}
\STATE{{$p \leftarrow RandomSelect(p^n)$}}
\STATE{{$d,a \leftarrow HumanExtract(p)$}}
\STATE{{$p_{de} \leftarrow Desensitization(p)$}}
\STATE{{$d' \leftarrow \mathbb{I}(p_{de},a,D)$}}
\STATE{{$\varepsilon \leftarrow \varepsilon+\mathbb{E}[\frac{\|d \cap d'\|}{\|d\|}]$}}
\ENDFOR
\RETURN{$\varepsilon/T$}
\end{algorithmic}
\end{algorithm}

As illustrated in Algorithm \ref{idenlink}, given a user prompt dataset $p^n$ and an identifier inference algorithm $\mathbb{I}$. Initially, direct identifiers for each prompt are extracted from the dataset to form a vector $D$, $\|D\|$ is set to 5 in this paper. Subsequently, a prompt is randomly selected from the dataset, and its direct identifiers $d$ along with private attributes $a$ are obtained. Following this, the prompt undergoes desensitization to achieve a desensitized prompt $p_{de}$. The inference algorithm $\mathbb{I}$ is then employed to select direct identifiers $d'$ from D, and the inference results are subsequently assessed. This entire process is repeated $T$ times, and the average value $\varepsilon_{i}$ is returned.

\subsection{Usability metircs}
Anonymization methods inevitably lead to a loss of data utility. Effective anonymization methods strive to achieve lower utility loss and a high level of privacy protection. In evaluating the utility loss of prompts, it is important to consider prompt semantic loss, measures of usability in large model inference responses, and readability metrics.

\textbf{Semantic loss(SL).} 
 In the process of prompt desensitization, effectively measuring the semantic loss can help ensure that the model maintains an accurate grasp of task semantics even after desensitization. We using Sentence-BERT \cite{reimers2019sentencebert}, a pre-trained model. Based on the Siamese network structure, this model is trained by maximizing the loss of semantic similarity.

\textbf{Inference loss(IL).}
The most straightforward and effective measure of utility is to compare the inference results of the large model before and after desensitization. This can help determine whether desensitization has affected the model’s semantic understanding. We utilize cosine similarity for this purpose, as it is unaffected by dimensionality and is suitable for processing sparse vectors. The formula for calculating cosine similarity is as follows:

\begin{align*}
    Cosine\,Similarity(A,B) = \frac{A \cdot B}{|A| \times |B|}
\end{align*}

\textbf{Readability loss(RL).}
We use Perplexity (PPL) as the readability metric. By comparing the PPL of the large model’s inference results before and after desensitization, we can understand the impact of desensitization on the model’s generation results, particularly with regard to semantics and fluency. The formula for Perplexity (PPL) is as follows:
\begin{align*}
    PPL = 2^{-\frac{1}{N}\sum_{i=1}^{N} \log P(w_i)}  
\end{align*}

\section{Experiment evaluation}
\label{exp}

\subsection{Experiment setup}
\subsubsection{Implementation}
All programs in our experiments are implemented using Python language (version 3.8.18). The fine-tuning of the llama-2 model is conducted on various datasets based on PyTorch (version 1.13.1), Transformers (version 4.28.1), and PEFT (version 0.6.3).

\textit{CSP} used the GPT-3.5 API for our experiments. \textit{DePrompt} runs on a Ubuntu 22.04 system server equipped with a V5000 GPU with 24GB of VRAM, alongside a 6-core processor and 24GB of RAM.

\subsubsection{Datasets}
\label{datasets}

In our experiments, the following datasets and models were utilized:

\textbf{HC3} is a corpus of Human-ChatGPT comparisons that aims to investigate how close ChatGPT is to Human Experts. To this end, it collect about questions from various public question answering datasets (e.g., medicine, law, finance QA) and the corresponding human answers and ChatGPT answers. The English samples in HC3 contain 24K questions, 59K human answers, and 27K Chatgpt answers \cite{H3C}.

\textbf{AlpacaGPT4} is deemed as an optimized version of Alpaca dataset, it includes 52K instruction tracking data samples for LLM fine-tuning. Compared to the original Alpaca dataset, AlpacaGPT4 also uses text-davinci-003 to complete the prompts, but generating the completions with GPT-4. Thus, the responses are of higher quality and lenght \cite{AlpacaGPT4}.

\textbf{UltraChat} is an open-source, large-scale, multi round dialogue data project supported by Turbo APIs, aimed at promoting the construction of powerful language models with universal dialogue capabilities. It consists of three parts: Questions about the World, Writing and Creation, and Assistance on Existing Materials, including 1.4M examples of technology, art, entrepreneurship, and writing \cite{UltraChat}.

\textbf{Swype-instruct} is a combination of multiple sources, including the GPT4All dataset, the Alpaca dataset from Stanford, custom generation using AllenAI augmentation, and some dataset augmentation from open-source Meta datasets. The dataset contains 0.88M instruction data for training and evaluating language models on various tasks \cite{swype}.

\subsection{Named Entity Recognition}
\subsubsection{Scene recognition}
From the aforementioned datasets in section \ref{datasets} , we collected 1000 data entries for each of the three scenarios: medical, everyday, and financial. Each set of test data was recorded and categorized as correct or incorrect, denoting accurate or inaccurate predictions, respectively. Subsequently, the prediction accuracy for each scenario and the overall accuracy were computed.

\begin{figure}[h]
\centering
\subfigure[\label{fig:scene1}]{\includegraphics[scale=0.3]{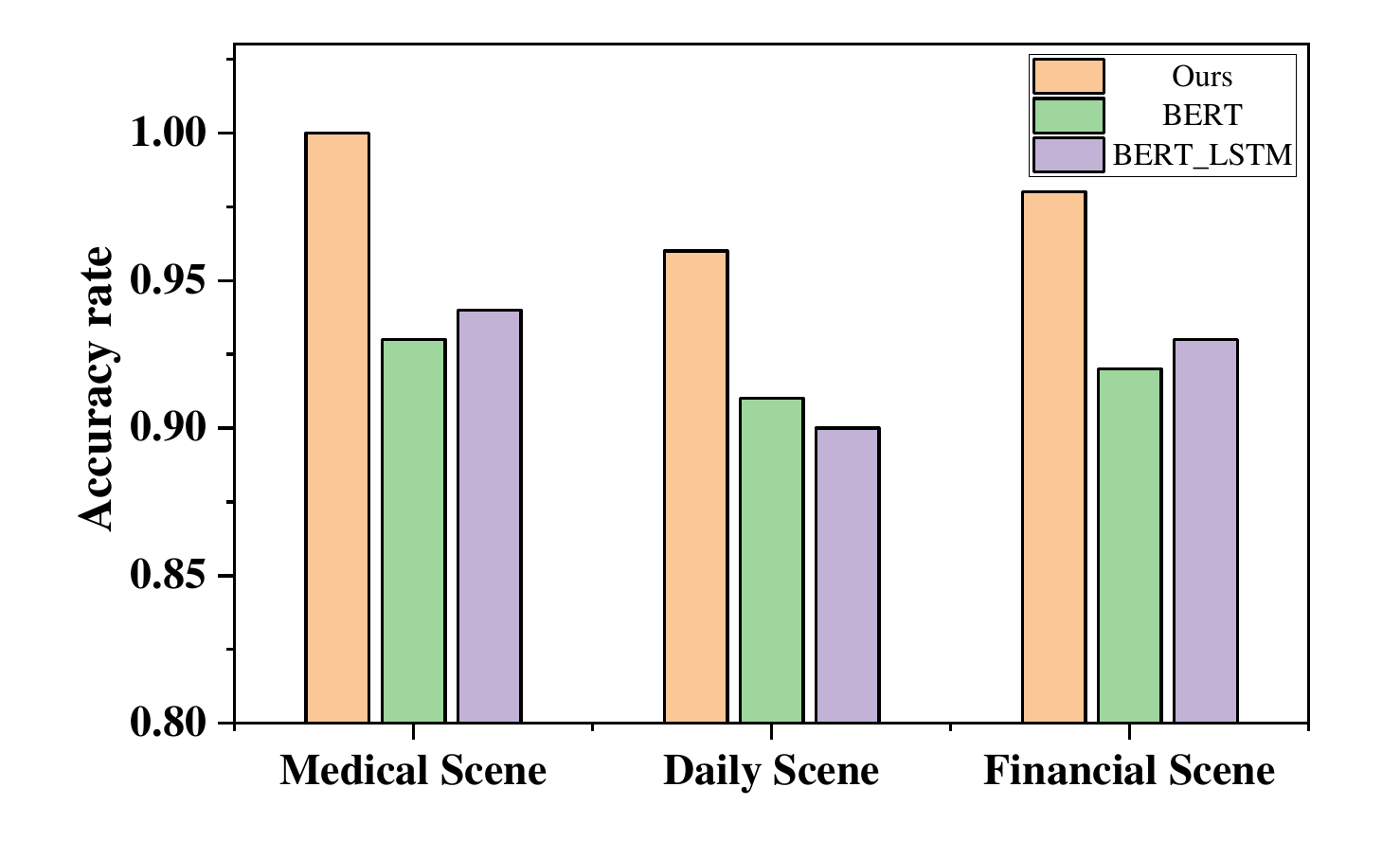}}
\subfigure[\label{fig:scene2}]{\includegraphics[scale=0.3]{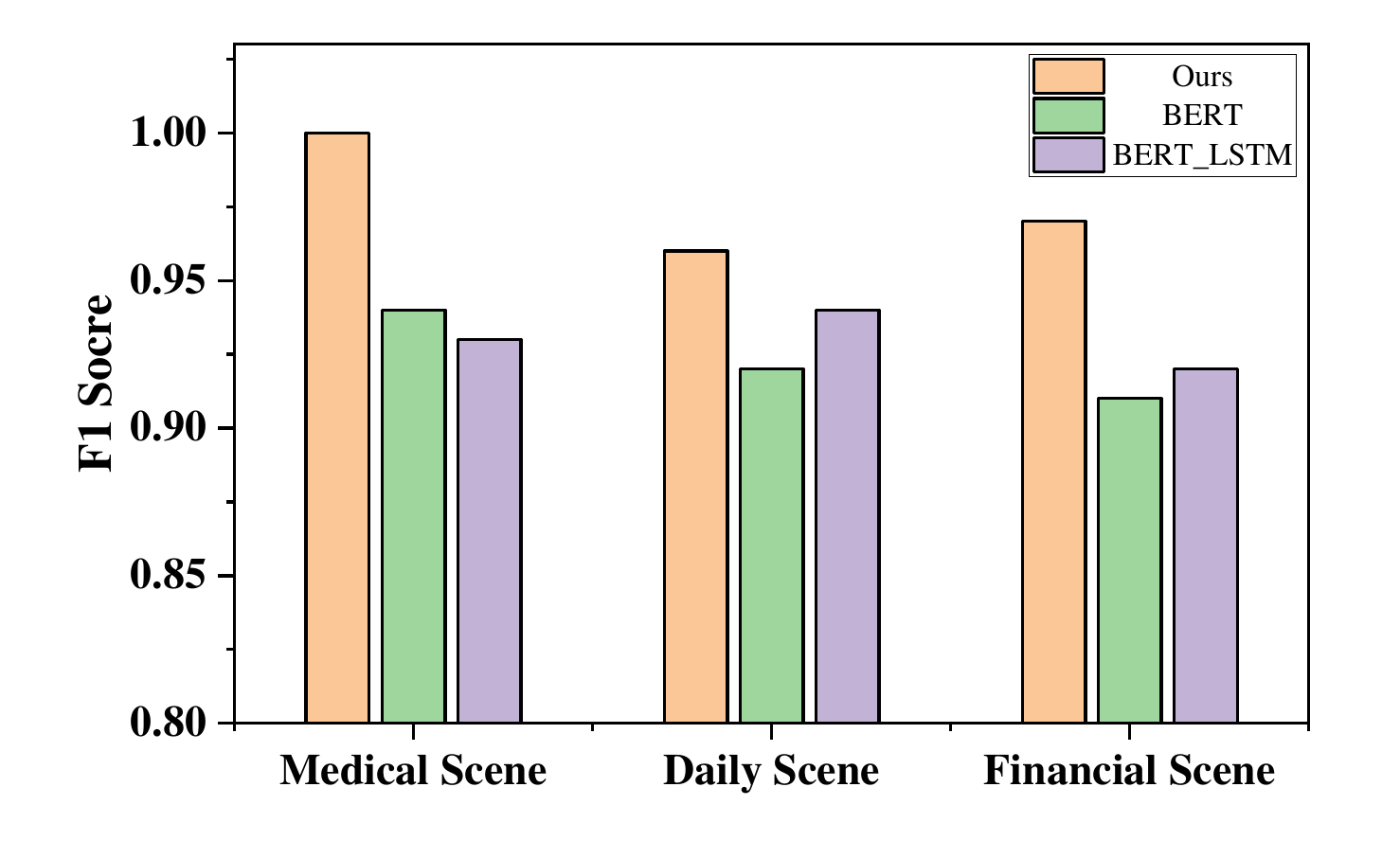}}
\\
\caption{Comparison of accuracy in scene recognition. (a) the Accuracy rate (b) the F1 score.}
\label{fig:scene}
\end{figure}

\figurename~\ref{fig:scene} compares our scheme with the BERT \cite{DBLP:journals/corr/bert} and BERT\_LSTM \cite{pandey2023bertlstm} models, revealing that our scheme demonstrates superior accuracy in scene recognition tasks. The overall accuracy of our model in scene recognition is 98\%. Notably, the highest predictive accuracy is achieved in the medical domain at 100\%, attributed to the prevalence of specialized medical terminologies in this context. However, the recognition accuracy for everyday scenes is relatively lower, owing to the complexity and broad scope of textual data within this category.

\subsubsection{Private entity extraction}
\label{PIIextractexp}
In this experiment, our aim is to validate the efficacy of our approach in privacy entity recognition tasks. Initially, we assess the accuracy of our approach in defining privacy in section \ref{extraction} and compare it with the BERT\_CRF model and the BERT\_LSTM\_CRF model \cite{liu2023automatic}. Subsequently, we proceed to perform experimental testing using the benchmark dataset \cite{pilan2022textbenchmark}. 

We collected 1400 instances from the datasets in section \ref{datasets},  ensuring a minimum of 200 instances for each privacy entity type, and annotated them manually.  As indicated in the \tablename~\ref{accuracyextract}, our approach outperforms the BERT and BERT\_LSTM models in the accuracy of privacy entity recognition. The overall accuracy of our privacy entity recognition model’s inference is 96\%. The predicted accuracy for medical funding and payment privacy data is relatively low, possibly due to a scarcity of medical payment data in the dataset, which the model did not learn well.

\begin{table*}[t]
\center
\small

\begin{tabular}{l|ccccccc|c}
\hline
 Models & PERSON & CODE & CONTACT & HEALTH & MEDICAL & PAYMENT & ASSET & AVERAGE \\ \hline
 BERT\_CRF & 89.78 & 88.72 & 90.91 & 75.32 & 81.98 & 75.45 & 83.32 & 83.64 \\
 BERT\_LSTM\_CRF & 91.43 & 90.32 & 94.11 & 81.33 & 89.56 & 79.32 & 90.25 & 88.05\\
 Llama3 & 98.81 & 97.32 & 98.17 &91.43& 91.46 & 90.31 & 96.87 & 94.91\\
 Qwen2 & 91.21 & 96.78 & 98.34 & 91.48 & 91.46 & 89.91 & 96.15 & 93.62\\
 GPT-3.5-turbo & 98.69 & 97.21 & 99.45 & 91.98 & 91.78 & 90.13 & 98.08 & 95.33\\
 DePrompt(Ours) & \textbf{98.88} & \textbf{98.73} & \textbf{99.56} & \textbf{92.23} & \textbf{92.56} & \textbf{91.13} & \textbf{98.56} & \textbf{95.95}\\ \hline
\end{tabular}
\caption{{\label{accuracyextract}The accuracy of private entity extraction.}}
\end{table*}

\begin{figure}[h]
\centering
\subfigure[\label{fig:benchmark1}]{\includegraphics[scale=0.3]{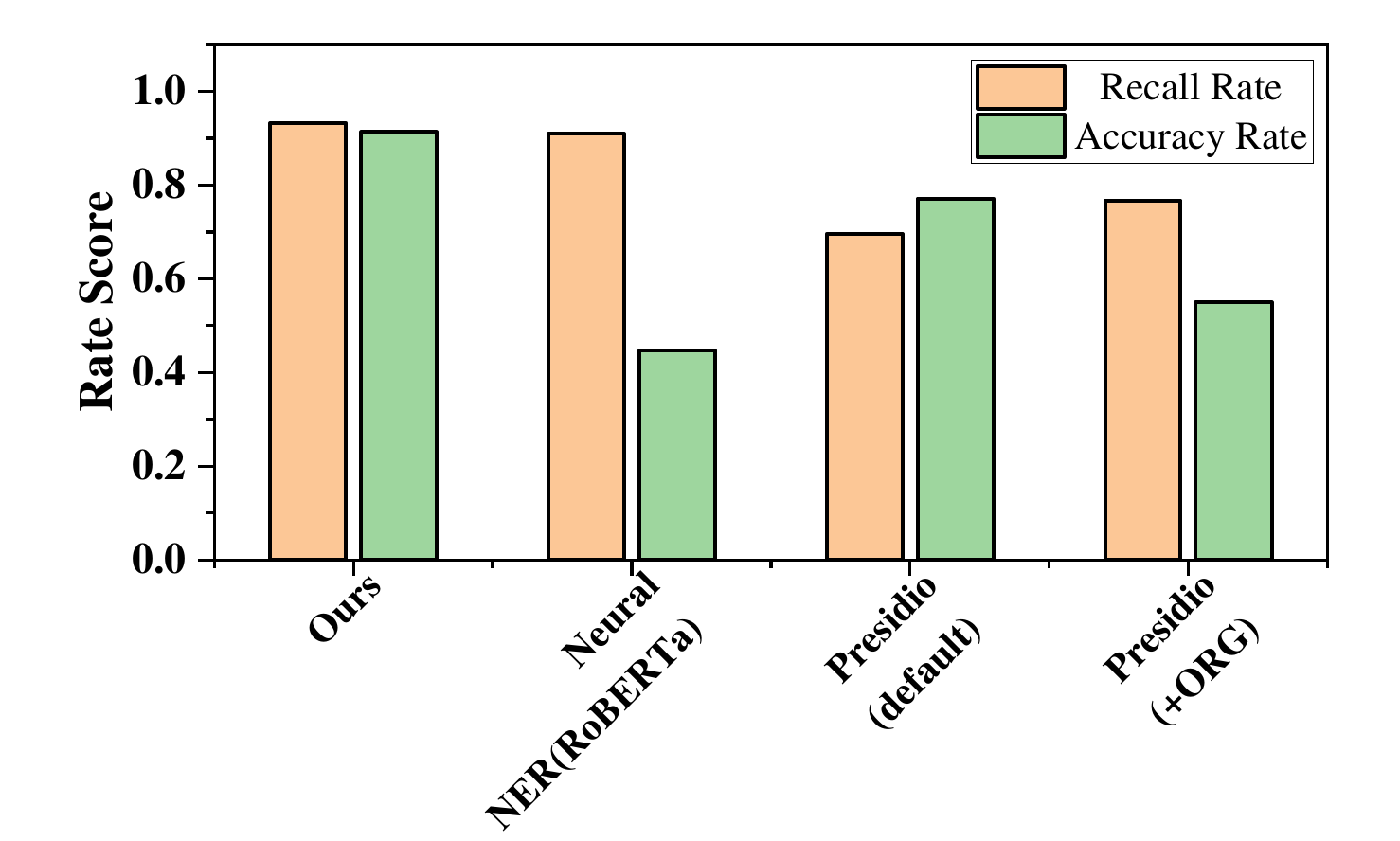}}
\subfigure[\label{fig:benchmark2}]{\includegraphics[scale=0.3]{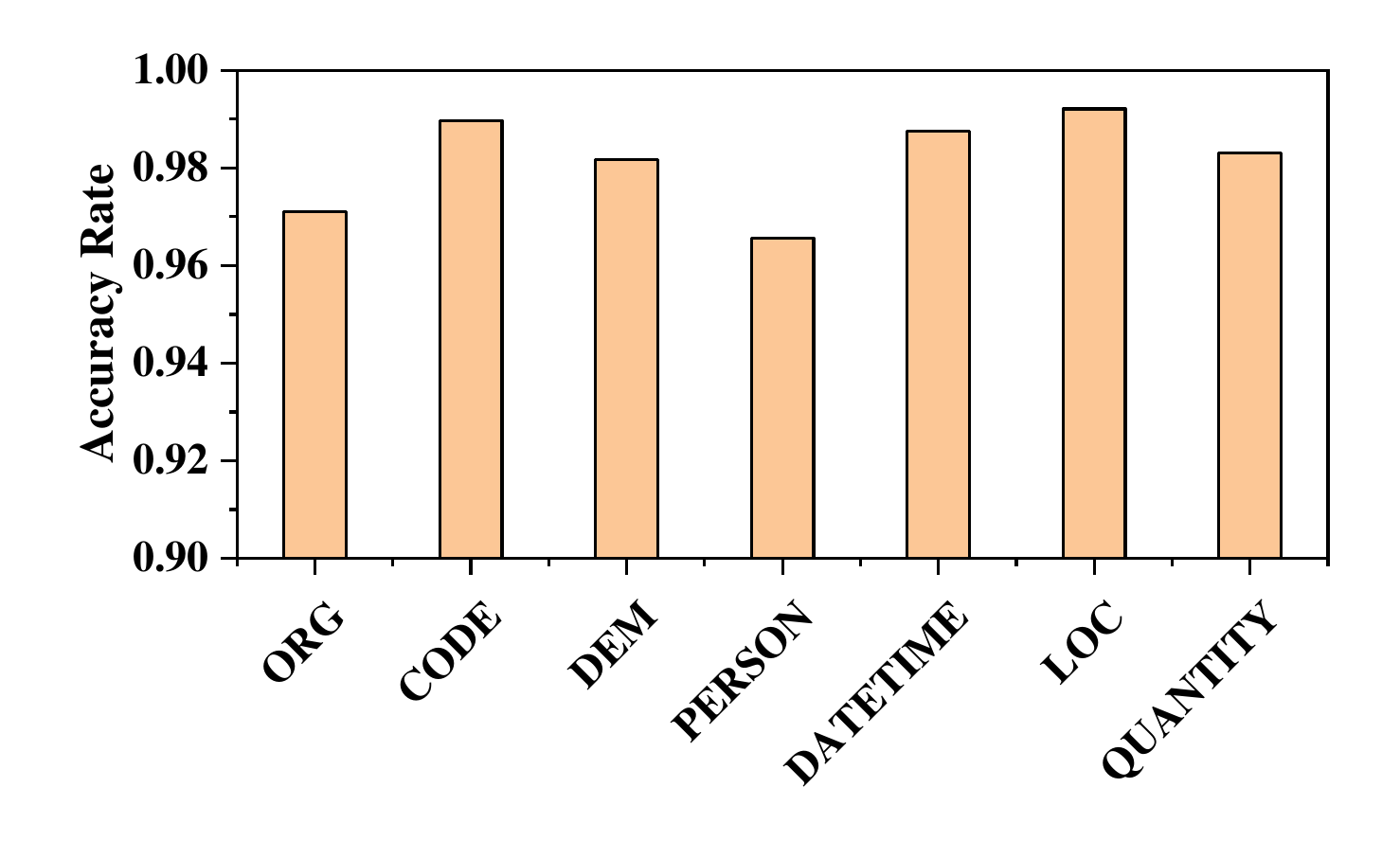}}
\\
\caption{\textbf{(a)}: Comparison of private entity extraction solutions in terms of the precision and recall; \textbf{(b)}: The accuracy of the benchmark dataset's privacy entities.}
\label{fig:benchmark}
\end{figure}

We downloaded the benchmark dataset for experimental comparison, focusing on three methods outlined in the benchmark: Neural NER (RoBERTa), Presidio (default), and Presidio (+ORG), with specific results shown in \figurename~\ref{fig:benchmark1}. From the figure, it is evident that our approach exhibits a certain advantage in both accuracy and recall, demonstrating its effectiveness on the benchmark dataset. We also analyzed the predicted accuracy of various privacy entities as defined in the benchmark, as depicted in \figurename~\ref{fig:benchmark2}.

\subsection{Privacy and Utility Evaluation}
\subsubsection{Overall Performance}
\begin{figure}
\centering
\includegraphics[width = \columnwidth]{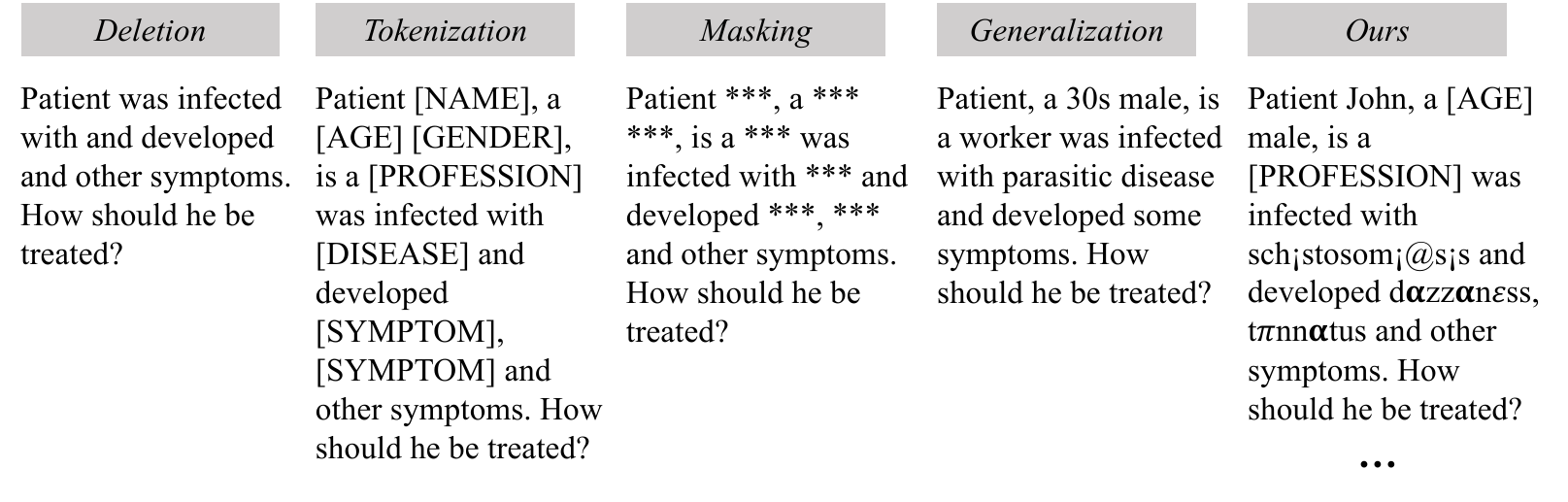}
\caption{Examples of various anonymization methods.}
\label{fig:afterde}
\end{figure}

In this subsection, we compare our schemes with traditional anonymization methods, including Deletion, Tokenization, Masking, and Generalization, using the privacy and utility metrics defined in section \ref{evalu}. Examples of prompts after anonymization are shown in the \figurename~\ref{fig:afterde}. We utilize the data described in the preceding section \ref{PIIextractexp} and process the prompts using various anonymization methods before uploading them to the GPT-3.5 API to obtain inference results.

\begin{table}
\renewcommand\arraystretch{1.3}
\fontsize{8}{9}\selectfont{}
\center
\begin{tabular}{lccccc}
\hline
 Methods & $\varepsilon_e$ & $\varepsilon_i$ & $SL$ &  $IL$ & $RL$  \\ \hline
 Deletion & \textbf{0.046} & \textbf{0.21} & 0.76  & 0.18 & 12.04 \\
 Tokenization & 0.052 & 0.22 & 0.88  & 0.22 & 7.64  \\ 
 Masking & 0.049 & \textbf{0.21} & 0.72 & 0.17 & 9.37\\
 Generalization & 0.124 & 0.67 & 0.82 & 0.64 &  4.28\\
 Ours & 0.097 & 0.24 & \textbf{0.92}  & \textbf{0.96} & \textbf{2.30} \\\hline
\end{tabular}
\caption{{\label{comparision}Evaluation and comparison of various desensitization methods.}}
\end{table}

As illustrated in the \tablename~\ref{comparision}, we set the relevant hyperparameters of our scheme to K=3 and N=5 for comparison with other methods. In terms of privacy assessment, Deletion yielded the optimal impact, while Generalization displayed the poorest performance. This discrepancy arises from the direct removal of PII in Deletion, contrasting with Generalization’s retention of privacy attributes. Although our method did not achieve the best privacy protection effect, our adversarial generation approach can thwart an attacker’s ability to ascertain the veracity of attributes within the privacy vector. Moreover, our method effectively disrupts the linkage between identifiers and attributes, thereby further reducing privacy risk.

In terms of prompt usability, our approach has achieved state-of-the-art results. Despite Deletion and Masking yielding the best privacy protection effects, the desensitized prompts generated by these methods generally lack usability, a similar issue observed with the Tokenization approach. While the Generalization method has achieved a certain balance between privacy and usability, it falls slightly short in both aspects compared to our approach. Therefore, these results show can be argued that our approach achieves the best current balance between privacy and usability.

\section{Conclusion}
\label{conclusion}
In this paper, we investigate the tradeoff between privacy and utility in Prompt-based data anonymization. Specifically, our study focuses on three fundamental characteristics—semanticity, linkability, and uncertainty. By leveraging fine-tuned LLM, we propose an adversarial generative desensitization approach for anonymizing the prompt. Additionally, we introduce an anonymization and evaluation framework specifically designed for prompts. Our experimental results unequivocally demonstrate the superiority of our proposed framework in terms of entity recognition performance and its ability to strike a favorable balance between privacy and utility. Moving forward, expanding this framework to a wider range of contexts where semantic understanding is crucial presents a compelling avenue for future exploration.
% \clearpage
\bibliography{reference}

\end{document}